\documentclass[pre,twocolumn,showpacs,preprintnumbers,amsmath,amssymb]{revtex4}
\usepackage{graphicx}% Include figure files
\usepackage{dcolumn}% Align table columns on decimal point
\usepackage{bm}% bold math
\usepackage{color}
\begin{document}

%\preprint{APS/123-QED}

\title{Deterministic excitable media under Poisson drive: power law
responses, spiral waves and dynamic range}

\author{Tiago L. Ribeiro}%
\email{tlr@df.ufpe.br}
\author{Mauro Copelli}%
 \email{mcopelli@df.ufpe.br}
\thanks{corresponding author}
%\homepage{http://www.df.ufpe.br/\~{}mcopelli}
\affiliation{%
Laborat\'orio de F{\'\i}sica Te\'orica e Computacional, Departamento
de F{\'\i}sica, Universidade Federal de Pernambuco, 50670-901 Recife, PE, Brazil}%

%\date{\today}% It is always \today, today,
             %  but any date may be explicitly specified

%(Received 2 July 2007; revised manuscript received 11 December 2007)

\begin{abstract}
When each site of a spatially extended excitable medium is
independently driven by a Poisson stimulus with rate $h$, the
interplay between creation and annihilation of excitable waves leads
to an average activity $F$. It has recently been suggested that in
the low-stimulus regime ($h \sim 0$) the response function $F(h)$ of
hypercubic deterministic systems behaves as a power law, $F \sim
h^m$. Moreover, the response exponent $m$ has been predicted to
depend only on the dimensionality $d$ of the lattice, $m = 1/(1+d)$
[T. Ohta and T. Yoshimura, Physica D {\bf 205}, 189 (2005)]. In
order to test this prediction, we study the response function of
excitable lattices modeled by either coupled Morris-Lecar equations
or Greenberg-Hastings cellular automata. We show that the prediction
is verified in our model systems for $d=1$, 2, and 3, provided that
a minimum set of conditions is satisfied. Under these conditions,
the dynamic range --- which measures the range of stimulus
intensities that can be coded by the network activity --- increases
with the dimensionality $d$ of the network. The power law scenario
breaks down, however, if the system can exhibit self-sustained
activity (spiral waves). In this case, we recover a scenario that is
common to probabilistic excitable media: as a function of the
conductance coupling $G$ among the excitable elements, the dynamic
range is maximized precisely at the critical value $G_c$ above which
self-sustained activity becomes stable.  We discuss the implications
of these results in the context of neural coding.
\end{abstract}

% PACS, the Physics and Astronomy
                             % Classification Scheme.
%\keywords{Suggested keywords}%Use showkeys class option if keyword
                              %display desired
\maketitle

\section{Introduction}

Sensory stimuli impinge continuously onto the peripheral nervous
system, where they are transduced into electrical activity of sensory
neurons. Understanding how those and subsequent neurons encode and
process the information of the stimulus remains a formidable challenge
for neuroscience since the pioneering work of Adrian~\cite{Adrian26},
and is the subject of ongoing research (see, e.g.,
Ref.~\cite{Bhandawat07} for recent progress on olfaction).

One of the most remarkable achievements of the nervous systems of
multicellular organisms is their large dynamic range, i.e., their
ability to cope with stimulus intensities which vary by {\em many\/}
orders of magnitude. Experimental evidence in this direction is
abundant, the simplest example being the century-old psychophysical
laws: the psychological perception $F$ of a given stimulus intensity
$h$ has been shown to be a power law for weak stimuli, $F \sim h^m$.
This behavior of the response curve $F(h)$ is known as Stevens' law,
and the response exponent $m$ is called Stevens' exponent in the
psychophysical literature~\cite{Stevens}. Microscopic (i.e., neural)
data also confirm this scenario: the activity of relay stages in
sensory processing also increases as a power law of the stimulus
intensities (e.g. glomeruli and mitral cells for
olfaction~\cite{Friedrich97,Wachowiak01}, or ganglion cells of the
retina~\cite{Deans02,Furtado06}). In both cases (psychophysical and
neural), the response exponents are typically less than $1$, which
indicates (as we will see below) a large dynamic range of the
response curves.

That large dynamic ranges should be evolutionarily favorable is
generally agreed upon, owing to the fact that natural stimuli indeed
span several decades of intensity. However, experimental results
show that the dynamic range of the very first sensory neurons which
perform the initial transduction is usually small, their firing rate
varying essentially linearly with stimulus intensity (see, e.g.,
Ref.\cite{Rospars00} for the case of olfaction). Therefore, what
remains to be explained is how those apparently conflicting results
can be reconciled. In other words, how can large dynamic ranges be
implemented by neurons?

Two main mechanisms have long been proposed. The first one is
adaptation, by which neurons manage to adjust their range of
operation according to the statistics of the ambient
stimulus~\cite{Normann74,Werblin74a,Werblin74b,Kim03,Borst05}. The
second one is the intrinsic variation of firing thresholds among a
population of sensory neurons, which would allow them to cover a
wide range of stimuli (in spite of each of them having a small
range)~\cite{Cleland99}. Both mechanisms can indeed contribute to an
enhancement of dynamic range. However, note that neither adaptation
nor threshold variation requires interactions among neurons to work,
insofar as adaptation has been understood as a dynamical process
which neurons undergo individually and the firing threshold of a
sensory neuron in principle does not depend on the activity of other
sensory neurons. Therefore, if these were the only mechanisms
responsible for enhancement of sensitivity and dynamic range, there
should be no significant change in those properties if lateral
connections among neurons were blocked.

Experimental data, however, suggest otherwise. Deans et
al.~\cite{Deans02} have measured the response function (firing rate
vs light intensity) of retinal ganglion cells of mice. For wild-type
mice, they found a class of cells that responded with large dynamic
range. When the same experiment was repeated with connexin36
knockout mice (i.e., mice that lack electrical synapses), they found
that both sensitivity and dynamic range were significantly reduced.
This suggests a third mechanism for dynamic range enhancement, based
on the interaction among neurons.

This third mechanism is the subject of the present contribution.
Previous work has revealed that, when excitable neurons are coupled
(via chemical or electrical synapses), the response function of the
resulting excitable medium indeed has much enhanced sensitivity and
dynamic range~\cite{Copelli02, Copelli05a, Copelli05b, Kinouchi06a,
Furtado06, Copelli07, Wu07, Assis08}, as compared to those of
isolated neurons. The underlying mechanism relies on very general
properties of excitable media: incoming stimuli generate excitable
waves which will disappear (due to the nonlinearity of their
dynamics) upon collision with one another and/or with the system
boundaries. For weak stimuli, waves are rare and can propagate a
long way before annihilation, therefore amplification is large (as
compared with what would be observed for uncoupled neurons); for
strong stimuli, waves contribute little to the overall network
activity (since most neurons are being externally driven), therefore
amplification is small. As a result, the medium as a whole has much
larger sensitivity and enhanced dynamic range as compared to those
of its building blocks~\cite{Copelli02, Copelli05a, Copelli05b,
Kinouchi06a, Furtado06, Copelli07, Wu07, Assis08}.

The above reasoning has been tested and confirmed in a variety of
models. In Refs.~\cite{Kinouchi06a, Copelli07, Wu07, Assis08} the
coupling among excitable elements was {\em probabilistic\/} (say,
via a transmission rate $\lambda$). In such a scenario, low-stimulus
amplification as described above occurs via {\em stochastic\/}
excitable waves, whose (finite) lifetimes are essentially
proportional to $\lambda$ (for small $\lambda$). The dynamic range
then initially increases with increasing $\lambda$, up to a critical
value $\lambda=\lambda_c$, where the system undergoes a
nonequilibrium phase transition. Above $\lambda_c$ self-sustained
activity becomes stable (i.e., small fluctuations can lead to
non-zero density of active sites even in the absence of external
stimuli). This hinders the coding of weak stimuli (just as a whisper
cannot be heard in a sound system dominated by audio feedback), a
problem that only worsens if the coupling is further increased. The
dynamic range then decreases above $\lambda_c$ and one concludes
that the maximum dynamic range is obtained precisely at the phase
transition~\cite{Kinouchi06a}.

Due to their probabilistic nature, the above cited systems were cast
in a framework of stochastic lattice models, from which useful
insights could be obtained by applying mean field approximations and
relying on well-known results of the statistical physics of
nonequilibrium phase transitions. For instance, the response
exponent $m$ at criticality was shown to be a critical
exponent~\cite{Kinouchi06a,Copelli07,Wu07,Assis08} whose value has
been known for over two decades~\cite{Marro99}. This should be
contrasted with the models employed in Refs.~\cite{Copelli02,
Copelli05a, Copelli05b}, where the coupling among excitable elements
was {\em deterministic\/}. In these papers, the models were such
that no self-sustained activity was observed for vanishing stimulus
rates. Besides, even if a transition to the self-sustained regime
occurred, the standard results from statistical physics would not be
easily applicable due to the deterministic nature of the excitable
waves.

In this context, our aim here is to fill two gaps: first, we verify
the existence of power law responses in deterministic excitable
media without self-sustained activity; second, we probe the
robustness of these power laws. To accomplish the first goal, we
have chosen to simulate hypercubic excitable media. This allowed us
to test a theoretical prediction which has recently been proposed
(based on scaling arguments) for the dependence of the response
exponent $m$ on the dimensionality $d$~\cite{Ohta05}. Moreover, it
reveals important differences (regarding the dependence of $m$ on
$d$) with systems where coupling is probabilistic (as recently
studied~\cite{Assis08}). To accomplish the second goal, we employed
the same model to show that, with a small change in its parameters,
self-sustained activity can occur, thus setting limits on the
validity of the theoretical prediction. As it turns out, this last
result puts the deterministic and probabilistic cases in a similar
state of affairs, where the dynamic range is maximized precisely at
the transition to self-sustained activity.

This paper is organized as follows. In Sec.~\ref{models}, the two
models employed are described. The response functions in the absence
and presence of self-sustained activity are analyzed in
Secs.~\ref{powerlaws} and~\ref{spiral}, respectively. From these
response functions we obtain the dynamic range, which is dealt with
in Sec.~\ref{dynamicrange}. Our conclusions are summarized in
Sec.~\ref{conclusions}.

\section{\label{models}Models}

In our simulations, we make use of a lattice in which each excitable
site $i$ is governed by the Morris-Lecar (ML)
equations~\cite{Morris-Lecar,Rinzel98b}

\begin{eqnarray}
C_m \dot V_i &=& - I^{ion}_i(V_i,w_i) +
I^{syn}_i\left(V_i,\{V_j\}\right) \nonumber \\
&& + I^{stim}_i(t)\; , \\
\dot w_i &=& \phi\left[ w_{\infty}(V_i) - w_i\right] \cosh\left(\frac{V_i - 10}{29}\right) \; ,\\
I^{ion}_i (V_i,w_i) &=& G_{Ca}m_{Ca}(V_i)(V_i - E_{Ca}) + G_{K}w_i(V_i - E_{K}) \nonumber \\
&& + G_{m}(V_i - V_{rest})\; , \\
m_{Ca}(V_i) &=& 0.5\left[ 1 + \tanh\left(\frac{V_i +
    1}{15}\right)\right]\; ,\\
w_{\infty}(V_i) &=& 0.5\left[ 1 + \tanh\left(\frac{V_i -
    10}{14.5}\right)\right]\; ,
\end{eqnarray}
where the membrane capacitance per unit area is $C_m=1\mu$F/cm$^2$,
membrane voltages $V_i$ are measured in mV, current densities in
$\mu$A/cm$^2$, $\phi = 1/3$~ms$^{-1}$, and maximal conductances for
calcium, potassium, and passive membrane leakage are respectively
$G_{Ca}=1$~mS/cm$^2$, $G_{K}=2$~mS/cm$^2$, and
$G_{m}=0.5$~mS/cm$^2$. The corresponding reversal potentials are
$E_{Ca}=100$~mV, $E_{K}=-70$~mV, and $V_{rest}=-35$~mV. Note that
the gating variable for calcium $m_{Ca}$ is assumed to be always in
equilibrium, while $w$ (which gates potassium currents) obeys a
first-order dynamics~\cite{Rinzel98b} (both are dimensionless). All
times are expressed in milliseconds.

Even though the ML equations were developed originally to describe
the membrane potential of the barnacle muscle fiber, our aim here is
not to model any specific biological tissue in particular, but
rather to shed light on the influence of the network topology on its
response properties, particularly the dynamic range. Here we study
hypercubic lattices with dimensionality $d$, restricting ourselves
to the simplest case of electrical coupling, for which the synaptic
currents are given by Ohm's law,

\begin{equation}
I^{syn}_i\left(V_i,\{V_j\}\right) = \sum_{j}^{2d} G_{ij} \left(V_j -
V_i\right)\; ,
\end{equation}
where $j$ runs over the first neighbors of $i$. The conductance
$G_{ij}$ between sites $i$ and $j$ could account for gap junctions
(e.g. as observed in axoaxonal contacts in the
hippocampus~\cite{Traub99b,lewis00} or dendrodendritic contacts of
mitral cells in the olfactory glomeruli~\cite{Kosaka05}) or ephaptic
interactions (as modeled by Bokil et al. to occur in the olfactory
nerve~\cite{Bokil01}).

The external current $I^{stim}_i(t)$ accounts for the stimuli arriving
in the network, which we model as a Poisson process. Each neuron
independently receives current pulses at constant rate $h$ (measured
in ms$^{-1}$). Each pulse has duration $D$ and intensity $I_0$ (so
that for $h \gtrsim D^{-1}$ the regime of a continuous external
current is approached).

To test the robustness of the results and to allow for larger system
sizes, we also simulate lattices in which each excitable element is
modeled by the $n$-state deterministic Greenberg-Hastings cellular
automaton (GHCA)~\cite{Greenberg78}. In this case, each site $i$ at
discrete time $t$ can be in states $x_i(t) \in \{0,1,2,\ldots,n-1
\}$, where $x=0$ and $1$ represent a quiescent (polarized) and
spiking (depolarized) neuron, respectively, whereas for $2 \leq x
\leq n-1$ the site is refractory. The dynamical rules are cyclical:
if $x_i(t)\geq 1$, then $x_i(t+1) = [x_i(t) + 1] \mod n$, i.e.,
after a spike the model neuron deterministically undergoes $n-2$
refractory steps before returning to the $x=0$ quiescent state. If
$x_i(t)=0$, then $x_i(t+1)=1$ if at least one of its $2d$ nearest
neighbors is spiking at time $t$ or if an external stimulus arrives
at site $i$ [$x_i(t+1)=0$ otherwise]. The Poissonian external
stimulus occurs independently at each site with probability $P = 1 -
\exp(-h\delta)$, where $\delta=1$~ms is the time step adopted in
this case.

For both models, $i=1,\ldots,N$, where $N = L^d$ is the total number of
excitable elements in a network of linear size $L$.

\section{\label{response}Response of hypercubic excitable media}

Let $F$ be the mean firing rate, defined as the total number of spikes
in an interval $T_{max}$, divided by the number $N$ of neurons and by
$T_{max}$. To avoid undersampling in the low-stimulus regime, we have
chosen $T_{max} = \max \{\bar{n}/(hN) ,100~\mbox{ms}\}$, where
$\bar{n}$ is the approximate mean number of attempts to initiate an
excitable wave (we have typically employed $\bar{n} = 25$). We define
the response function (or transfer function) of the network to the
external stimulus as $F(h)$. In the following, we make use of a
uniform coupling $G_{ij}=G$ and study how the response function
changes with $G$.

\subsection{\label{powerlaws}Power laws}

\begin{figure}[bht]
\centerline{\includegraphics[width=0.99\columnwidth]{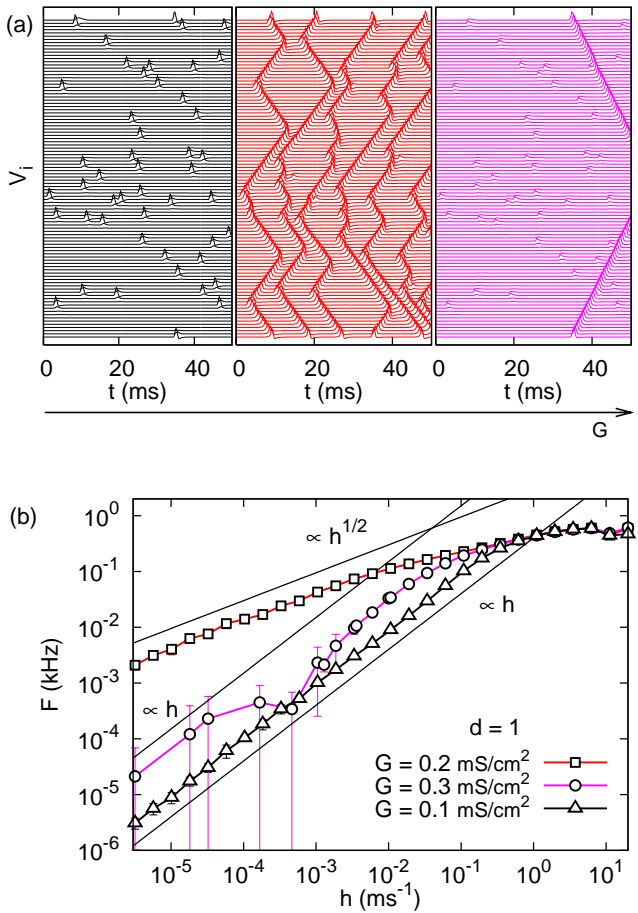}}
\caption{\label{fig:response1d} (Color online) (a) Membrane
potentials of 100 ML neurons versus time with $h=10^{-2}$~ms$^{-1}$
for $G=0.1$ (left panel), $0.2$, (middle panel) and $0.3$~mS/cm$^2$
(right panel). The seed of the pseudo-random number-generator is the
same for the three values of $G$. (b) Response function for a
one-dimensional lattice of $L=1000$~ML excitable elements. Symbols
(bars) represent averages (standard deviations) over 10 runs. Solid
lines are power laws discussed in the text.}
\end{figure}

Figure~\ref{fig:response1d} shows the results for a one-dimensional
ML lattice with $D = 0.3$~ms and $I_0 = 15$~$\mu$A/cm$^2$. As $G$
increases, three regimes are observed in the response of the
network. For weak coupling [left panel in
Fig.~\ref{fig:response1d}a, triangles in
Fig.~\ref{fig:response1d}b], synaptic currents from spiking
neighbors are not strong enough to generate spikes, so each stimulus
event generates one spike, and the response function increases
linearly (up to saturation at $F_{max}$, which is essentially the
inverse of the refractory period). Above a certain value
$G_1^\prime\simeq 0.14$~mS/cm$^2$, however, the conductance is
strong enough to allow the propagation of excitable waves. In this
regime [middle panel in Fig.~\ref{fig:response1d}a, squares in
Fig.~\ref{fig:response1d}b], which is observed up to a second
transition at $G_1^{\prime\prime} \simeq 0.24$~mS/cm$^2$, excitable
waves are created by external stimuli and annihilated by one another
and by the boundaries (open boundary conditions have been employed
throughout this paper). Above $G_1^{\prime\prime}$, current leakage
to neighbors is so large that it typically prevents neurons from
spiking upon the incidence of a single stimulus pulse. What we
observe [right panel of Fig.~\ref{fig:response1d}a] is that a neuron
will fire only if it is at the boundary (in which case it has fewer
neighbors and consequently less leakage) or if two stimulus pulses
happen to arrive nearly consecutively (in a mimicry of temporal
summation). Note that in the three panels in
Fig.~\ref{fig:response1d}a the seed of the pseudo-random-number
generator is the same, so the spikes in the left panel coincide with
stimulus pulses. In the right panel of Fig.~\ref{fig:response1d}a,
however, only the stimulus pulses that happened to fall right at the
borders generated waves (all other visible perturbations are
subthreshold, not spikes).  In this regime, inevitably poor
statistics ensues, except for large stimulus rates, as reflected in
the $G=0.3$ (circles) curve in Fig.~\ref{fig:response1d}b. Note,
however, that if a spike is finally produced, propagation of an
excitable wave does occur, which explains why the response in this
case is larger than for $G < G_1^\prime$.

\begin{figure}[!htb]
\centerline{\includegraphics[width=0.99\columnwidth]{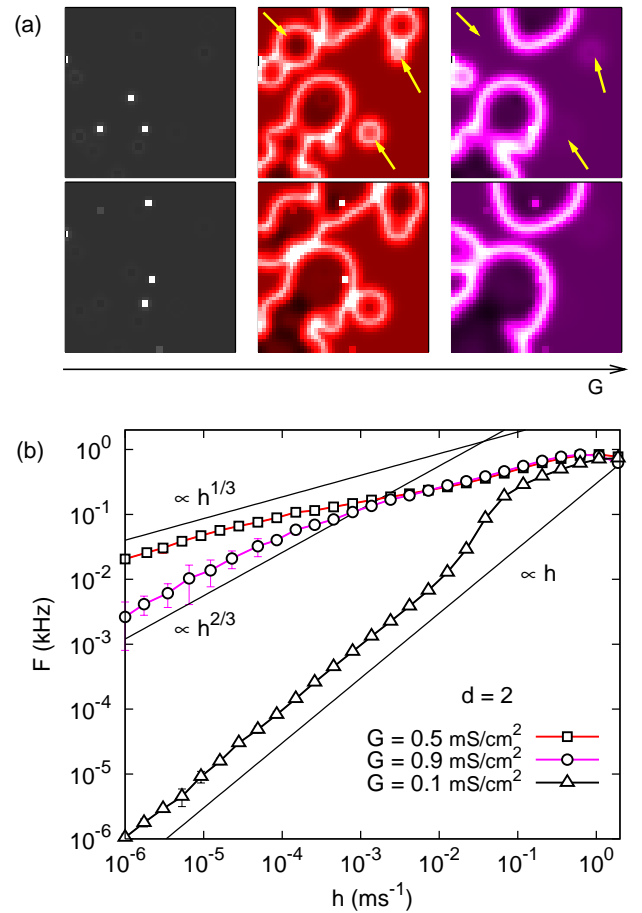}}
\caption{\label{fig:response2d} (Color online) (a) Snapshots of
networks with $50\times 50$ ML neurons, with depolarized (spiking)
membrane potentials coded as white. From left to right, $G=0.1$,
$0.5$, and $0.9$~mS/cm$^2$, with $h=2\times 10^{-3}$~ms$^{-1}$ and
$t=3.0$~ms ($3.5$~ms) for top (bottom) row. The seed of the
pseudo-random-number generator is the same for the three values of
$G$. (b) Response function for a network of $200\times 200$~ML
excitable elements. Symbols (bars) represent averages (standard
deviations) over ten runs. Solid lines are power laws discussed in
the text.}
\end{figure}

The response curves in Fig.~\ref{fig:response1d}b clearly show power
laws $F \sim h^m$ in the low-stimulus regime. For $G<G_1^\prime$,
the response is linear ($m=1$) and can be easily explained: for each
stimulus pulse, a small number of spikes is generated (typically
one) and excitable waves do not interact. For $G_1^\prime < G
<G_1^{\prime\prime}$, however, excitable waves are created in
randomly located points and annihilate upon encountering one
another. To understand how this nonlinear interaction leads to a
power law in the dependence of $F$ on $h$, Ohta and Yoshimura have
recently proposed an elegant scaling reasoning~\cite{Ohta05}. In the
scaling regime, $F$ should depend on a dimensionless variable $A$.
Since $h$ is small, the characteristic times for wave creation and
wave annihilation are much smaller than the time of free
propagation. Therefore, the only relevant parameters are the width
$l$ of an excitation, the wave speed $c$, and the rate $h$.
Recalling that $h$ is measured in events per unit time per site
(thus having dimension of $t^{-1}L^{-d}$), we obtain $A =
hc^{-1}l^{1+d}$. If we now assume a scaling relation $F \sim A^m$,
the exponent $m$ can be obtained by noting that in the low-stimulus
regime waves are sparsely distributed and the dependence of $F$ on
$l$ must be linear; hence $m=1/(1+d)$~\cite{Ohta05}.

As shown in Fig.~\ref{fig:response1d}b, this prediction is confirmed
in our one-dimensional ML simulations in the parameter region where
excitable waves propagate ballistically. Particularly for $d=1$, the
scaling relation $F \sim h^{1/2}$ had already been conclusively
confirmed for the GHCA (in both
simulations~\cite{Copelli02,Furtado06} and analytical
calculations~\cite{Furtado06}) and coupled map
lattices~\cite{Copelli05a}. However, for more realistic models, it
was only approximately verified for a chain of Hodgkin-Huxley model
neurons~\cite{Copelli02} and a reaction-diffusion partial
differential equation~\cite{Ohta05}, with exponents around $m \simeq
0.4$. In Fig.~\ref{fig:response1d}b we fill this gap with an
agreement over more than two decades.

In two dimensions, simulations have been carried out with stronger
stimulus pulses ($D=0.45$~ms and $I_0=150$~$\mu$A/cm$^2$) to prevent
excessive leakage owing to the larger number of neighbors. The same
scenario has been observed. For small $G$, each stimulus pulse
generates at most an evanescent wave with a radius of a few
neighbors [left panel of Fig.~\ref{fig:response2d}a]. For $G >
G_2^\prime \simeq 0.225$~mS/cm$^2$, however, generated waves can
propagate ballistically with their radii increasing indefinitely. As
shown in the middle panel of Fig.~\ref{fig:response2d}a,
annihilation in this case is more complicated than for $d=1$, for
now colliding waves may have different radii and their surfaces
merge to form irregular-shaped
excitations~\cite{lewis00,Copelli05a,Copelli05b}. This regime breaks
down for $G = G_2^{\prime\prime} \simeq 0.725$~mS/cm$^2$, above
which current leakage is again too strong and spikes are generated
with at least two nearly consecutive stimulus pulses or at the
boundaries. Note that several waves that appear in the middle panel
of Fig.~\ref{fig:response2d}a are absent in the right panel, as
exemplified by the arrows (as in Fig.~\ref{fig:response1d}, the seed
is the same for the three panels). Several waves in the right panel
of Fig.~\ref{fig:response2d}a have been created at the borders (and
propagate faster than those of the middle panel because $G$ is
larger).

As for the response functions, Fig.~\ref{fig:response2d}b shows that
Ohta and Yoshimura's exponent $m=1/3$ for $d=2$ agrees (for two
decades) with simulations for $G=0.5$~mS/cm$^2$ (and this holds true
in the whole interval $G_2^\prime < G <
G_2^{\prime\prime}$). Interestingly, another exponent (not predicted
originally~\cite{Ohta05}) arises for $G>G_2^{\prime\prime}$: in this
regime, waves typically require two nearly consecutive stimulus pulses
to be created, and for weak stimuli this occurs approximately at a
rate $h^\prime \sim h^2$. But once they are created, Ohta and
Yoshimura's reasoning is still valid, now with the dimensionless
variable rewritten as $A = h^\prime c^{-1}l^{1+d}$. We therefore
obtain the exponent $m=2/(1+d)$, which is reasonably confirmed for
$G=0.9$~mS/cm$^2$ (circles) in Fig.~\ref{fig:response2d}b. Looking
back to the analogous situation for the one-dimensional case, the
circles in Fig.~\ref{fig:response1d}b are compatible with an exponent
$m=1$ (the extremely poor statistics notwithstanding). Whether further
increasing $G$ leads to other transitions (inducing the necessity of,
say, $k>2$ nearly consecutive pulses to generate a wave) and new
exponents [presumably $m=k/(1+d)$] is a question beyond the scope of
this work, but perhaps worth pursuing. It is important to point out,
however, that these transitions may have limited biological
applicability: chemical synapses (which do not suffer from leakage)
are not included in this model, yet abound in the nervous system.

\begin{figure}[!htb]
\centerline{\includegraphics[width=0.99\columnwidth]{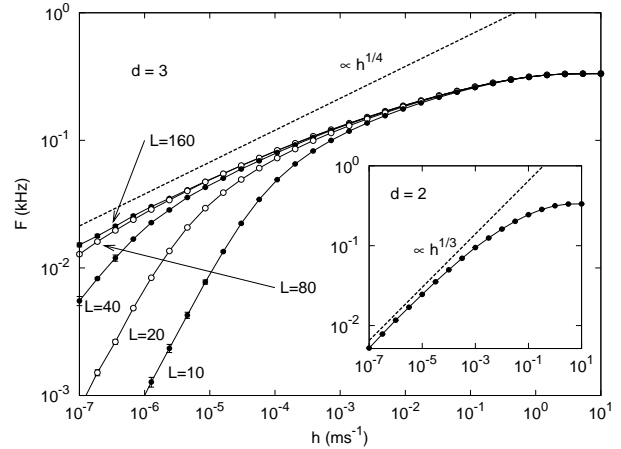}}
\caption{\label{fig:response3d} Response curves of the GHCA model in
$d=3$ for increasing system size ($n=3$).  Inset: GHCA response
function for $d=2$ ($L = 2744$, averages over five runs). Dashed
lines shows the response exponent $m=1/(1+d)$ for both cases.}
\end{figure}

In order to test Ohta and Yoshimura's prediction in three
dimensions, we have performed simulations of the GHCA model. With
the rules defined in Sec.~\ref{models}, an incoming stimulus pulse
generates an excitable wave which propagates ballistically until
annihilation with another wave or with the system
borders~\cite{Traub99b,lewis00,Copelli05b}, precisely as observed in
the intermediate region $G^\prime < G < G^{\prime\prime}$ for the ML
equations. The motivation for switching to a simpler model is that
it allowed us to simulate much larger networks than would be
feasible with the ML equations. As shown in the response functions
of Fig.~\ref{fig:response3d}, finite-size effects are strong.
However, for a network of $N = 160^3$ automata (a system size beyond
our computational resources for the ML equations), it is already
possible to verify the power law $F \sim h^{1/4}$ for more than two
decades. Incidentally, we note that the response function of
two-dimensional GHCA networks has been studied in
Ref.~\cite{Copelli05b}, but the power law has been missed. The inset
of Fig.~\ref{fig:response3d} confirms the predicted exponent.

\subsection{\label{spiral}Spiral waves}

\begin{figure}[!htb]
\centerline{\includegraphics[width=0.99\columnwidth]{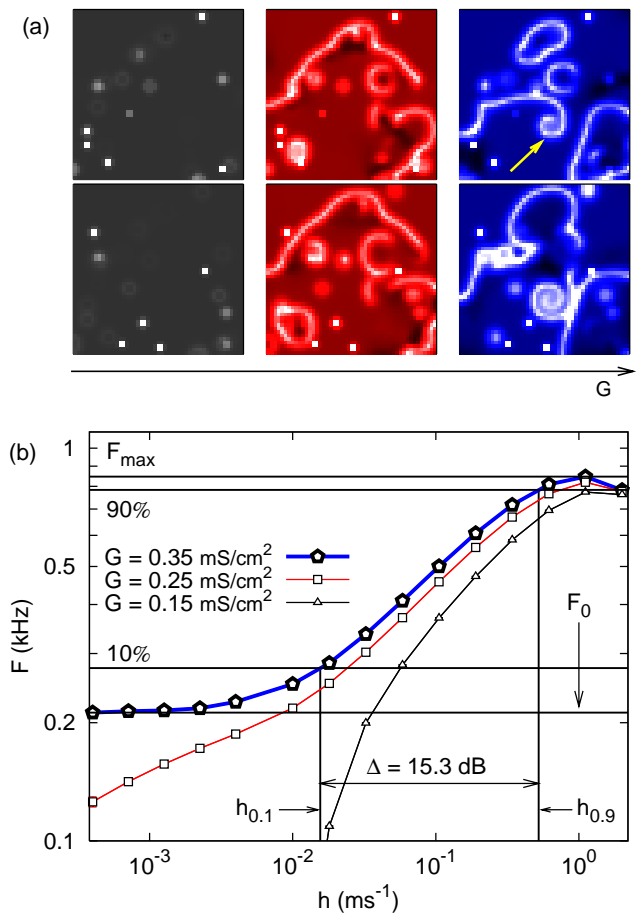}}
\caption{\label{fig:spiral} (Color online) Spiral waves with
  $\phi=0.4$~ms$^{-1}$. (a) Snapshots of networks with $50\times 50$ ML
  neurons, with depolarized (spiking) membrane potentials coded as
  white. From left to right, $G=0.15$, $0.25$, and $0.35$~mS/cm$^2$, with $h=4\times 10^{-3}$~ms$^{-1}$ and
  $t=20$~ms ($21$~ms) for top (bottom) row. The seed of the
  pseudo-random-number generator is the same for the three values of
  $G$. (b) Response function for a network of $200\times 200$~ML
  excitable elements. Symbols represent averages over five runs (standard
  deviations are smaller than symbol size). Firing rates were measured
  over 150~ms after a 50~ms transient. Horizontal and vertical lines
  illustrate the relevant quantities for calculating the dynamic range
  $\Delta$ (see Eq. (7) and text for details).}
\end{figure}

What we have described so far suggests that the response exponent is
indeed $m=1/(1+d)$ whenever the following two conditions are
satisfied: (A) every quiescent neuron (i.e., not only those at the
borders) spikes upon the incidence of a {\em single\/} stimulus
pulse and (B) this spike creates a deterministic excitable wave
which will be annihilated at the borders or upon encountering other
wave(s). In the examples shown in figures~\ref{fig:response1d}
and~\ref{fig:response2d}, these two conditions are simultaneously
satisfied only for $G^\prime < G < G^{\prime\prime}$. For $G <
G^\prime$, condition A is satisfied, but B is not; for $G >
G^{\prime\prime}$, condition B is satisfied, but A is not.

The above scenario, however, is not general. As has been known for
many decades, excitable media can exhibit self-sustained activity in
the form of spiral or scroll waves, a topic that has received much
attention due to its relevance in different scientific branches such
as cardiology~\cite{QuWG99,WeissCQKG00},
cytology~\cite{LechleiterGPC91}, physics~\cite{MerzhanovR99},
chemistry~\cite{Winfree72,PetrovOS97}, and
neuroscience~\cite{Karma93}, among others. In
Fig.~\ref{fig:response2d} the parameters of the ML system were such
that spiral waves did not occur. With a slight deviation in
parameter space, however, spiral waves may appear, even in a
homogeneous lattice. As shown in the right panel of
Fig.~\ref{fig:spiral}a, this is the case for $\phi=0.4$~ms$^{-1}$
and $G=0.35$~mS/cm$^2$ (all other parameters remaining the same),
for instance. In this system, spiral waves emerge [see arrow in
Fig.~\ref{fig:spiral}a] because of the local inhomogeneities created
by the stochastic input~\cite{Jung97} and, once established, they
typically resist being destroyed by the same stochastic input (even
though their shape is continuously perturbed by the Poisson pulses).

With this new phenomenon at play, how does the scenario evolve as
the coupling $G$ changes? For low $G$ (say, $G<G_2^\prime$), the
overall behavior of the system is the same as that of
Fig.~\ref{fig:response2d}, i.e., a stimulus-induced spike at one
site does not propagate too far [compare the left panels of
Figs.~\ref{fig:response2d}a and ~\ref{fig:spiral}a].
Correspondingly, the response function is linear. If $G$ is
increased, a transition occurs which allows the wave radii to
increase indefinitely [Fig.~\ref{fig:response2d}a and
Fig.~\ref{fig:spiral}a, middle panels]. However, contrary to what
was previously observed, this dynamic regime is no longer valid in a
broad range of $G$ values. As $G$ is further increased, spiral waves
quickly emerge. As for the second transition previously observed at
$G_2^{\prime\prime}$, it now essentially loses meaning, for as soon
as the waves are created
--- no matter whether by one or two incoming stimuli, or at the
boundaries --- the conditions are set for the spiral waves to
dominate the network.

Regarding Ohta and Yoshimura's conjecture in this scenario, the
response function near the transition to self-sustained activity
suffers from strong statistical fluctuations, as expected [see solid
squares in the inset of Fig.~\ref{fig:deltas}d]. It seems compatible
with a power law with exponent $m=1/3$, but for less than a decade
only [note that even the self-sustained activity suffers from
finite-size effects for low enough stimulus rate --- see pentagons
in the inset of Fig.~\ref{fig:deltas}d]. It is at present unclear
whether larger system sizes or longer stimulus times would confirm
the power law at the transition.

The drastic consequences of this self-sustained activity for the
response curve are shown in Fig.~\ref{fig:spiral}b: the
weak-stimulus response no longer decreases as a power law for
decreasing $h$, but reaches instead a value $F_0$ which corresponds
to the average firing rate when the lattice is dominated by spiral
waves. To obtain a reasonable estimate of $F_0$, we simulated the
following protocol: $150\times 150$ networks were stimulated during
a period $T_{stim}=100$~ms with a constant rate $h=4\times
10^{-3}$~ms$^{-1}$. The stimulus was then switched off ($h=0$) and
the mean activity of the network $F_0$ was measured after a
transient $T_{trans}=900$~ms. Figure~\ref{fig:deltas}c shows how
$F_0$ depends on $G$. A transition is clearly seen near $G\simeq
0.275$~mS/cm$^2$, above (below) which $F_0 \geq 0$ ($F_0=0$). In the
inset of Figure~\ref{fig:deltas}c we also show the probability $p$
that spiral waves survive after the transient, which was estimated
by dividing the number of runs in which spiral waves survived by the
total number of runs. The sharpness of the $p(G)$ curve also
suggests a transition to a regime where self-sustained activity is
stable.

This second scenario appears to be more general than the one
described in Sec.~\ref{powerlaws}. We have simulated networks in
which each element was modeled by the Hodgkin-Huxley
equations~\cite{HH52} with standard parameters~\cite{Koch} and have
obtained spiral waves. Moreover, one of the most studied causes of
spiral wave creation is disorder and noise in the excitable
dynamics~\cite{Jung98,GarciaOjalvo99b,Lindner04}, which are absent
from the present study. We have nonetheless tested some ML networks
where $\phi$ was distributed around $1/3$~ms$^{-1}$ with some
variance, and have again obtained spiral waves. It is important to
remark that, for the purposes of the present study, it is not enough
that an excitable medium be able to sustain spiral waves in the
absence of stimulus, say, for a given initial condition. The
question is whether the Poisson stimulus is able to create spiral
waves and, at the same time, allow them to survive. Consider, for
instance, the limit of very weak stimuli ($h\to 0$). In this regime
spiral waves hardly emerge (even for $\phi=0.4$~ms$^{-1}$) because
fluctuations are not sufficiently strong [see, e.g., the open
pentagons in the inset of Fig.~\ref{fig:deltas}d]. At the other
extreme, a large value of $h$ can easily provide the necessary
fluctuations, but then the created spiral waves will be
statistically overshadowed by the very stimuli that generated them.
Overall, the probability of self-sustained activity coexisting with
the Poisson stimulus depends not only on the model parameters (in
this case, $\phi$ or $G$) but also on the system size ($N$),
stimulus rate ($h$), and duration ($T_{max}$). A more detailed study
of this dependence would be welcome.

\section{\label{dynamicrange}Dynamic range}

\begin{figure*}[!htb]
%\centerline{\includegraphics[width=0.99\columnwidth]{Deltas.eps}}
\centerline{\includegraphics[width=0.8\textwidth]{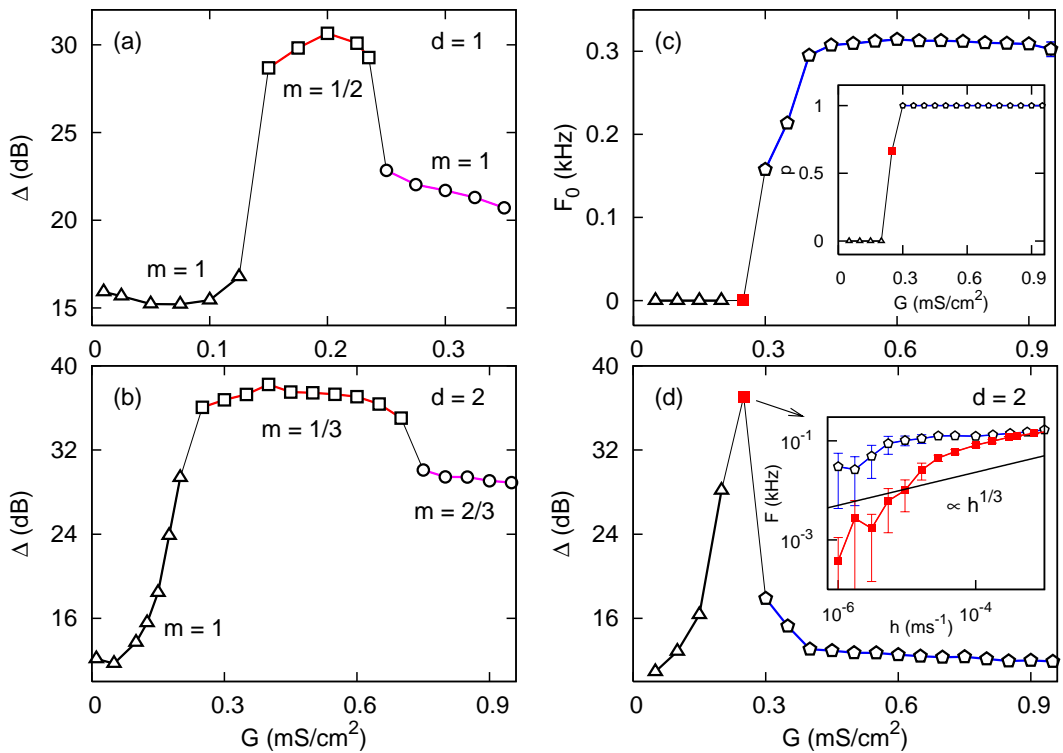}}
\caption{\label{fig:deltas} (Color online) Left (right) column:
  absence (emergence) of spiral waves. Dynamic range versus coupling
  conductance for $\phi=1/3$~ms$^{-1}$ [(a) and (b)] and
  $\phi=0.4$~ms$^{-1}$ (d). Triangles denote $G<G^\prime$ and squares
  denote $G^\prime < G < G^{\prime\prime}$.  Circles denote
  $G>G^{\prime\prime}$ in the absence of self-sustained activity (b),
  whereas pentagons denote the spiral wave regime (d). In (c), the
  self-sustained activity $F_0$ in the absence of stimulus is plotted
  against $G$ for fixed $\phi=0.4$~ms$^{-1}$. Inset of (c): estimated probability $p$ of spiral wave
  survival versus $G$ (see text for details). Inset of (d): response
  functions with $\phi=0.4$~ms$^{-1}$ for $G=0.25$ (solid
  squares) and $0.275$~mS/cm$^2$ (open pentagons). System sizes are
  $L=1000$ in (a); $N=200^2$ in (b) and (d); and $N=150^2$ in (c).}
\end{figure*}

We can now return to the quantity that originally motivated this
study. The dynamic range $\Delta$ of a response curve $F(h)$ is
formally defined as~\cite{Firestein93}
\begin{equation}
\Delta = 10\log_{10}\left( \frac{h_{0.9}}{h_{0.1}} \right)\; ,
\end{equation}
where $h_{0.1}$ ($h_{0.9}$) is the stimulus intensity such that the
difference $F-F_0$ is $10$~\% ($90$~\%) percent of the response
interval $F_{max}-F_0$. As depicted in Fig.~\ref{fig:spiral}b,
$\Delta$ measures (in decibels) the range of stimulus intensities
that can be ``appropriately'' coded by the mean firing rate of the
system, discarding intensities whose corresponding responses are too
close either to saturation ($h>h_{0.9}$) or to baseline
($h<h_{0.1}$). This measure of appropriateness is evidently
arbitrary, but standard in the biological literature and very
useful, since it is a dimensionless quantity that allows direct
comparison with experimental results.

Figure~\ref{fig:deltas} shows the behavior of the dynamic range
(estimated numerically from the response curves) as a function of
the coupling conductance. For $d=1$ [Fig.~\ref{fig:deltas}a],
$\Delta$ changes very little for $G < G_1^\prime$, staying in the
range of $16$~dB (which is comparable to experimental values of
isolated olfactory sensory neurons~\cite{Rospars00} and retinal
ganglion cells of connexin36 knockout
mice~\cite{Deans02,Furtado06}). The transition near $G_1^\prime$
seems abrupt, after which the dynamic range becomes substantially
larger: the system attains $\sim 31$~dB, an enhancement of about
$100$~\% which had also been previously obtained with a cellular
automaton model~\cite{Furtado06}. This enhancement is clearly due to
a change in the response exponent $m$, which greatly amplifies weak
stimuli [recall the squares in Fig.~\ref{fig:response1d}b]. For
$G>G_1^{\prime\prime}$ the dynamic range is reduced, once more
because of the change in the weak-stimulus sensitivity [recall the
circles in Fig.~\ref{fig:response1d}b]. It is important to remark
that the poor statistics in Fig.~\ref{fig:response1d}b do not
compromise the accuracy of the measured dynamic range, since the
strong fluctuations occur below the sensitivity threshold $h_{0.1}$.

For $\phi=1/3$~ms$^{-1}$, the results in $d=2$ are similar [see
Fig.~\ref{fig:deltas}b]. As $G_2^\prime$ is approached from below,
the transition is somewhat smoother than for $d=1$. More
importantly, since the response exponent for $G_2^\prime < G <
G_2^{\prime\prime}$ ($m=1/3$) is {\em smaller\/} than for the
corresponding regime in $d=1$ ($m=1/2$), the weak-stimulus
amplification for $d=2$ is {\em larger\/} and so is the dynamic
range, which reaches $\sim 38$~dB. The same trend in the dependence
of $\Delta$ on $d$ is observed in the GHCA model: with a fixed
system size $N=14^6$, by varying the dimensionality, we obtain
$\Delta=31$, $43$ and $54$~dB for $d=1$, 2, and 3, respectively.
Note that these values are comparable to those obtained in the ML
model (the differences in $d=2$ being explained by finite-size
effects, as extensively discussed in Ref.~\cite{Copelli05b}).

This picture changes qualitatively when spiral waves come into play
($\phi=0.4$~ms$^{-1}$, rightmost column of Fig.~\ref{fig:deltas}).
For $G<G_2^\prime$ the dynamic range increases monotonically with
$G$, reaching a maximum near $G_2^\prime$. Increasing $G$ further,
however, leads to the onset of spiral waves, and the nonzero
baseline activity $F_0$ prevents the appropriate coding of weak
stimuli. This is clearly seen in Fig.~\ref{fig:spiral}b (pentagons):
an observer would have much difficulty in distinguishing the
responses of any two points below $h=10^{-3}$~ms$^{-1}$, which leads
to a drastic decrease in dynamic range. Moreover this problem
becomes more and more severe as $G$ is further increased: since
$F_0$ increases with $G$ for $G>G_2^\prime$ [see
Fig.~\ref{fig:deltas}c], the dynamic range decreases with increasing
$G$. Therefore, if a deterministic excitable medium supports spiral
waves in some parameter region, its dynamic range will be maximum
precisely at the transition where they become stable.

%Finite size effects (linear response, see GHCA figura; delta increases
%logarithmically).

\section{\label{conclusions}Concluding remarks}

We have simulated hypercubic networks of excitable elements modeled
by the Morris-Lecar equations and Greenberg-Hasting cellular
automata. We have studied how the collective response $F$ of the
network to a Poisson stimulus with rate $h$ changes with the
coupling $G$ and the dimensionality $d$. Two scenarios have been
observed. In the first one, a broad range of $G$ values exists such
that excitable waves are created and thereafter propagate
ballistically, being annihilated upon encountering one another or
the system boundaries. In this regime, the response function
$F(h;d)$ is shown to be a power law $F \sim h^m$. Furthermore, we
have confirmed that, if waves are created upon the incidence of a
single stimulus pulse, the response exponent agrees with the
theoretical prediction of Ohta and Yoshimura,
$m=1/(1+d)$~\cite{Ohta05}. We have argued that in a regime where
wave creation requires the incidence of two nearly consecutive
stimuli, an exponent $m=2/(1+d)$ should be expected and is confirmed
by our ML simulations in $d=2$ (also for a broad range of $G$
values).

If a system is such that the exponent $m=1/(1+d)$ holds, the dynamic
range increases with the dimensionality $d$ (as confirmed here for
$d=1$ and 2 in the ML model and $d=1$, 2, and 3 for the GHCA model).
This is in stark contrast with probabilistic excitable systems,
where the maximum dynamic range attained at a given dimension $d$ is
a decreasing function of $d$. This happens because in that case $m$
corresponds to the critical exponent $\delta_h^{-1}$ (apparently
belonging to the directed percolation universality
class~\cite{Assis08}), and $\delta_h^{-1}$ increases with $d$.

In this context, one should not be misled by the apparent paradox
posed by the assumption that a deterministic system is ``just'' a
particular case of a probabilistic one. Consider, for instance, a
probabilistic version of the $d$-dimensional GHCA in which a
stimulus would be transmitted to its quiescent neighbors with
probability $q$: the function $\Delta(q)$ has qualitatively the same
shape as that of Fig.~\ref{fig:deltas}d and the {\em maximum\/}
value of $\Delta$ attained at given $d$ is a decreasing function of
$d$~\cite{Assis08}. Why then for $q=1$ do we have an increasing
$\Delta(d)$? Remember that the main condition for the exponent
$m=1/(1+d)$ to hold is the absence of self-sustained activity. In a
probabilistic system, this requires not only that $q$ is precisely
$1$, but also that the initial conditions are appropriately
set~\cite{Copelli05b,lewis00}. For $q$ infinitesimally smaller than
$1$ or $q=1$ with random initial conditions, self-sustained activity
ensues in the probabilistic GHCA. Therefore, in this particular
model the result $m=1/(1+d)$ is obtained only under very artificial
circumstances, at the edge of the parameter space and only for
restricted initial conditions. In contrast, for the deterministic ML
lattices studied here, the exponent holds in a broad region of the
parameter space for any initial condition.

A substantially different scenario has been obtained with a change
in a single parameter of the ML model, for which stable spiral waves
were observed when the coupling was increased above a certain
critical value (leading to a breakdown of Ohta and Yoshimura's
prediction). Given the ubiquity of spiral waves in studies of
excitable media, this scenario is likely to be more general than the
one previously described. In this case, a unifying picture emerges
for both deterministic and probabilistic excitable media: the
dynamic range in both cases is maximized at the critical value of
coupling above which self-sustained activity becomes stable.

Put into a broader context, our results reinforce the idea that
optimal information processing near criticality, a topic which has
received much attention in recent decades~\cite{Bak}, could have a
bearing on the brain sciences. In fact, experimental results that
are consistent with the hypothesis of neurons collectively operating
near a critical regime have recently
appeared~\cite{Beggs03,Beggs04,Eguiluz05,Plenz07}, joined by
theoretical efforts aimed at understanding the computations
themselves~\cite{BakChialvo03,Bertschinger04,Haldeman05,Kinouchi06a}
as well as the homeostatic mechanisms that could maintain the system
at criticality~\cite{deArcangelis06,Levina07}. These issues still
pose remarkable challenges for the years to come, which opens the
possibility of new lines of research connecting physicists with
systems biology in general, and neuroscience in particular.

\begin{acknowledgments}
T.L.R. and M.C. acknowledge financial support from Conselho Nacional
de Desenvolvimento Cient\'{\i}fico e Tecnol\'ogico (CNPq), CAPES,
PIBIC, and the special program PRONEX. The authors are also grateful
to O. Kinouchi for discussions and suggestions.
\end{acknowledgments}

\bibliography{copelli,spiralwaves.wos}% Produces the bibliography via BibTeX.

\end{document}